\documentclass[11pt,a4paper]{article} 
\usepackage{jcappub} 
\usepackage{epsfig}
\usepackage{epstopdf}
\usepackage{graphicx}
\usepackage{amsmath}
\usepackage{amssymb}
\usepackage{natbib}
\newcommand{\msun}{\ensuremath{M_{\odot}}}

\title{Understanding the Observed Evolution of the Galaxy Luminosity Function from $z=6$--$10$ in the Context of Hierarchical Structure Formation}

\author{
Joseph A.\ Mu{\~n}oz
}
\affiliation{	
Department of Physics and Astronomy, University of California Los Angeles\\
Los Angeles, CA 90095, USA
}
\emailAdd{jamunoz@astro.ucla.edu}

\abstract{
Recent observations of the Lyman-break galaxy (LBG) luminosity function (LF) from $z\approx6$--$10$ show a steep decline in abundance with increasing redshift.  However, the LF is a convolution of the mass function of dark matter halos (HMF)--which also declines sharply over this redshift range--and the galaxy-formation physics that maps halo mass to galaxy luminosity.  We consider the strong observed evolution in the LF from $z\approx6$--$10$ in this context and determine whether it can be explained solely by the behavior of the HMF.  From $z\approx6$--$8$, we find a residual change in the physics of galaxy formation corresponding to a $\sim 0.5$ dex increase in the average luminosity of a halo of fixed mass.  On the other hand, our analysis of recent LF measurements at $z \approx 10$ shows that the paucity of detected galaxies is consistent with almost no change in the average luminosity at fixed halo mass from $z \approx 8$.  The LF slope also constrains the variation about this mean such that the luminosity of galaxies hosted by halos of the same mass are all within about an order-of-magnitude of each other.  We show that these results are well-described by a simple model of galaxy formation in which cold-flow accretion is balanced by star formation and momentum-driven outflows.  If galaxy formation proceeds in halos with masses down to $10^{8}\,\msun$, then such a model predicts that LBGs at $z\approx10$ should be able to maintain an ionized intergalactic medium as long as the ratio of the clumping factor to the ionizing escape fraction is $C/f_{\rm esc}\lesssim10$.
}

\keywords{
galaxies: evolution -- galaxies: formation -- galaxies: high-redshift -- cosmology: observations -- cosmology: theory
}

\notoc
\begin{document}
\maketitle
\flushbottom

\section{Introduction}

Surveys with the WFC3 camera aboard the {\it{Hubble Space Telescope}} have now collected large samples of galaxies at redshifts $z>6$ using the Lyman-break technique \cite{Bouwens10b, McLure10, Bunker10, Yan10, Finkelstein10, Bouwens11b}.  Detections have even been made of J-band dropouts out to $z\sim10$ \cite[][hereafter B11]{Bouwens11a} with additional constraints placed on the abundance at brighter magnitudes \cite{Oesch11}.  The UV luminosity functions (LFs) extracted from these data clearly evolve significantly from $z\sim6$--$10$.  For example, B11 found one candidate J-dropout above their detection limit in the two-year Hubble Ultra Deep Field 09 dataset, yet nine were expected assuming no evolution in the LF from $z=8$.  Fitting an artificial Schechter function to the observed J-dropout LF yields $M^{\star}_{\rm UV}=-18.3$ when fixing a normalization and slope similar to those found from $z\sim4$--$6$ \cite{Bouwens07}.  Corresponding values determined at $z=7$ and 8 show a dimming $M^{\star}_{\rm UV}$ with increasing redshift.  However, the extent to which these values elucidate the physics of galaxy formation is unclear.

In the context of hierarchical structure formation, high-redshift Lyman-break galaxies (LBGs) are hosted by dark matter halos whose abundance drops rapidly from $z\sim6$--$10$, mirroring the behavior of the observed LF.  In theoretical models of the LF, the halo mass function (HMF) is convolved with a mapping from halo mass to galaxy luminosity that encodes the physics of galaxy formation.  We seek to determine whether any residual change is required in this physics to account for the observed changes in the LF or if its behavior can be understood entirely through that of the HMF.   Such a change is anticipated in a cold-flow model of galaxy formation \cite[e.g.,][]{Keres05,Dekel09a} where halos at higher redshift have higher growth rates than the same mass halos at lower redshift and so may be expected to be more luminous.  Fluctuations around the mean accretion rate also imply a range of galaxy luminosities for a given halo mass in contrast to many analytic models of the mass-to-light ratio \cite[e.g.,][]{SLE07, Trenti10}.

While hydrodynamical simulations of high-redshift galaxies \cite[e.g.,][]{Salvaterra11, Finlator11} can, in principle, answer these questions, they are more computational expensive and reliant on predetermined physical relations than are analytic models.  We use an analytic LF model that was initially developed by a companion paper \cite[][hereafter ML11]{ML11} to consider the minimum mass of halos that can host galaxies.  Here, however, we include the most recent LF data and improve upon our fitting method as we focus on the redshift evolution of the galaxy mass-to-light ratio.  In \S\ref{sec:model}, we briefly review this simple model and its free parameters.  We describe the fitting procedure in \S\ref{sec:fit} and consider the change in the average luminosity for fixed halo mass out to $z\approx10$ in \S\ref{sec:evol}.  \S\ref{sec:sigL} explores fluctuations around this mean luminosity, while \S\ref{sec:lc} investigates the effects of observing along the light-cone given the determined redshift evolution of the mass-to-light ratio.  We then compare our results with theoretical predictions of galaxy fueling and recent numerical studies in \S\ref{sec:galfuel} and \S\ref{sec:compare}, respectively.  Finally, in \S\ref{sec:discussion}, we discuss implications for future observations.

\section{Modeling the luminosity function}\label{sec:model}

The ML11 model for the high-redshift galaxy luminosity function has three components: (1) the Sheth-Tormen HMF \cite{ST02}, (2) a cutoff halo mass below which galaxies cannot form, and (3) a distribution function for the luminosity of a galaxy hosted in a halo of a given mass.  Since the HMF is well-understood, we briefly summarize the second two parts of the model below and refer the reader to ML11 for further details.

We assume a redshift-invariant critical suppression mass $M_{\rm supp}$ is required for a halo to form a galaxy.  This threshold is a free parameter that contains the physics of feedback processes, such as supernovae and photoionization, that prevent gas from condensing into lower mass halos.  However, since halos are built up from those of smaller mass, we expect some suppression of galaxy formation in halos more massive than $M_{\rm supp}$ as well.  Using a merger-tree algorithm based on the excursion-set formalism to quantify this effect, ML11 found that the ``active fraction" $\epsilon_{\rm AF}$, i.e. the probability that a halo of a given mass $M_{\rm h}$ can host a galaxy, is redshift-independent and well-approximated by:
\begin{equation}\label{eq:eAF}
{\rm log_{10}}\,\epsilon_{\rm AF}(M_{\rm h}, M_{\rm supp})=-(2.23\,{\rm log_{10}}[M_{\rm h}/M_{\rm supp}])^{-2.6}.
\end{equation}
The active fraction is distinct from the duty cycle, which is the fraction of an object's lifetime that it is active enough to be observable and controls the overall normalization of the LF.  By contrast, the active fraction we define here, uniquely specified by the ratio of halo mass to the free parameter $M_{\rm supp}$, is the fraction of halos with galaxies that have {\emph{ever}} been active and describes the faint-end cutoff in the LF.  For a fixed fraction of gas turned into stars per dynamical time, an instantaneous burst of star formation gives an exponentially declining SFR as a function of time.  Therefore, if a halo is counted in the active fraction, if it has ever been forming stars, then it will ever after have non-zero SFR and luminosity.

For these halos large enough to be actively forming stars, the luminosity emitted by the hosted galaxy is a log-normal distribution with mean $L_{\rm c}$ and standard deviation $\sigma_{\rm L}$:
\begin{equation}\label{eq:LDF}
\frac{dP}{d{\rm log_{10}}L}=\frac{1}{\sqrt{2\,\pi\,\sigma_{\rm L}^2}}\,{\rm exp}\left(-\frac{[{\rm log_{10}}(L/L_{\rm c})]^2}{2\,\sigma_{\rm L}^2}\right),
\end{equation}
where 
\begin{equation}\label{eq:L10}
L_{\rm c}=L_{10}\,\left(\frac{M_{\rm h}}{10^{10}\,\msun}\right)
\end{equation}
is proportional to halo mass.  The shape of equation \ref{eq:LDF} differs from the one assumed by Ref. \cite{Lee09}, who adopted a linear Gaussian profile, but is consistent with the conditional luminosity function approach elsewhere in the literature \cite[e.g.,][]{CO06, Bouwens08a}.  This distribution results from an excursion-set analysis of possible merger histories for halos of a given mass.  For example, a galaxy that has had a recent merger (or, equivalently, a rapid influx of gas) will be brighter than one whose last major burst of star formation was further in the past.  Our treatment obviates the need for a traditional duty cycle; if a halo is massive enough it will be forming stars.  Galaxies that have used up much of their gas in star formation and are too faint to be seen at the time of observation are simply in the tail of their luminosity distribution.  Instead of correcting our LFs for observability by eliminating faint galaxies from the calculated LF, as has been typical in previous studies with an occupancy or duty cycle approach \cite[e.g.,][]{Lee09}, these sources simply contribute to fainter luminosity bins beyond the detection threshold.  This interpretation is consistent with the results of numerical simulations \cite{Finlator11}.  Thus, our method replaces the common degeneracy between mass-to-light ratio and duty with a reliance on $\sigma_{\rm L}$, which describes the mixing of galaxies with different halo masses among luminosity bins.  In this picture of the luminosity distribution function, $\epsilon_{\rm AF}$ is effectively the fraction of halos lying in a second peak to the distribution with zero luminosity.  Assuming a simple model for the star formation in a disk with a constant fraction of gas turned into stars per dynamical time, a Monte Carlo analysis of halo merger trees finds a roughly constant value of $\sigma_{\rm L} \approx 0.25$.  In \S\ref{sec:sigL}, we further discuss the physical significance of $\sigma_{\rm L}$ and consider whether the data can constrain it {\it{a priori}}.  Since the $L_{\rm c}$ is proportional to halo mass, the average mass-to-light ratio, $\Upsilon=10^{10}\,\msun/L_{10}$, is independent of halo mass at a given redshift and uniquely specified by $L_{10}$ at that redshift.  $L_{10}$ is, therefore, the second free parameter of our model.  

For a given halo mass, equations \ref{eq:eAF} and \ref{eq:LDF} give the fraction of halos forming stars and the distribution of resulting luminosities, respectively.  However, if an unaccounted for feedback process further shuts off star formation in some fraction of halos, the resulting effective ``duty-cycle" will turn our fits of $L_{10}$ in \S\ref{sec:evol} into underestimates.  

This model gives the average abundance of galaxies in the universe at a given redshift.  However, there are two geometrical effects that can conspire with structure formation to make the observed LF deviate from the average.  First, the narrow field-of-view and large halo bias at high redshift can result in significant cosmic variance \cite{Somerville04, TS08, Munoz10}.  In the $z \sim 10$ case of B11, the fluctuation in abundance is of order $40\%$--large but still smaller than other uncertainties in the problem.  Additionally, the large redshift range probed by high-redshift dropout surveys, combined with the rapid evolution of the LF, can have profound implications for the galaxy sample, for example, a distribution toward lower redshifts, a higher abundance, and a flatter observed LF than would be expected if every galaxy in the sample were at the mean redshift \cite[][see \S\ref{sec:lc}]{ML08b}.  We show that the contribution of this effect to observations of the $z \sim 10$ LF is much smaller than the uncertainties of the measurements for our model in section \S\ref{sec:lc} and ignore it in the rest of this work. 

\section{The fitting method}\label{sec:fit}

We fit our model parameters to the data using a $\chi^2$ minimization method.  Focussing on the $L_{10}$, we marginalize over $M_{\rm supp}$, allowing the suppression mass to vary as we search for the value of $L_{10}$ that best fits the data.  Thus, our constraints also reflect uncertainties in $M_{\rm supp}$.  

Furthermore, we improve upon the ML11 constraints by including the upper limits in the LFs at $z\sim7$ and $\sim8$ in our analysis.  For each set of model parameters, magnitude bins of the LF with measured abundance $n_{\rm obs}$ and error $\sigma$ contribute an amount $d\chi^2=\left[\left(n_{\rm mod}-n_{\rm obs}\right)/\sigma\right]^2$ to the total value of $\chi^2$, where $n_{\rm mod}$ is the abundance predicted by the model for that set of parameters.  Additionally, each magnitude bin in which no galaxies were observed and having a $1\sigma$ upper limit of $n_{\rm up}$ contributes $d\chi^2=\left(n_{\rm mod}\,V-n_{\rm obs}\,V\right)^2/(n_{\rm mod}\,V)=n_{\rm mod}/n_{\rm up}$, where $V$ is the volume probed by the dropout search in the magnitude bin, and we estimate $V\approx1/n_{\rm up}$.

At $z\sim10$, B11 finds only one galaxy candidate detection (actually an estimate of $\sim 0.8$).  Therefore, in the associated magnitude bin, $\sigma$ does not include the full Poisson error around $n_{\rm mod}$ for sets of model parameters that produce $n_{\rm mod}<n_{\rm obs}$.  Instead, we modify the above procedure for this data point so that the error is due to Poisson variance about $n_{\rm mod}$ with an additional $50\%$ contribution from cosmic variance.

\section{The evolution of the luminosity function}\label{sec:evol}

In ML11, we presented results for the best-fit values for $M_{\rm supp}$ at $z\approx6$, 7, and 8 and found that all were consistent with each other at about $10^{9.4}\,\msun$.  While we were able to constrain $M_{\rm supp}$ strongly from above, a value of $10^{8}\,\msun$ could not be ruled out, and the best-fit values should be taken as tentative.  Here, we provide improved constraints for $L_{10}$ and extend the analysis to $z\approx10$.  In effect, we extract from the drop in galaxy abundance the contribution owing to the rapidly falling HMF and determine whether a complementary change is required in $L_{10}$, the parameter that describes the average mass-to-light ratio.  

We consider the observed LF of $z\sim6$ i-dropouts \cite{Bouwens07}, $z\sim7$ z-dropouts, $z \sim 8$ Y-dropouts \cite{Bouwens11b}, and J-dropouts at $z\sim10$ \cite{Oesch11}.  Further improving on the analysis of ML11, we use the LFs at $z\sim7$ and 8 from the final version of \cite{Bouwens11b}, which includes the full two-year HUDF09 results.  We assume a flat prior on $M_{\rm supp}$ for values greater than $10^{8}\,\msun$ and fit for $L_{10}$ by minimizing $\chi^2$ calculated as described in \S\ref{sec:fit}.  We find values of ${\rm log_{10}}\,L_{10} = 27.19^{+0.05}_{-0.06}$\footnote{Since the best-fit combination of $L_{10}$ and $M_{\rm supp}$ is barely ruled out at the $70\%$ level, the errors here represent the $95\%$ confidence interval.}, $27.36^{+0.11}_{-0.14}$, $27.69^{+0.14}_{-0.16}$, and $27.45^{+0.49}_{-0.33}$ at $z = 5.9$, 6.8, 8.0, and 10.2, respectively.  
These constraints are consistent with the preliminary estimates set in ML11.  The rise in $L_{10}$ from $z\sim6$--8 represents an average increase in luminosity at fixed halo mass by about 0.24 dex per redshift interval.  Because of the fitting to the observed LF, the values determined here represent observed emission after any extinction by dust.  Ref. \cite{Bouwens07} estimates a $\sim0.18$ dex dust correction at $z\approx6$, implying an intrinsic value of ${\rm log_{10}}\,L_{10} = 27.37$ at this redshift.  On the other hand, the blue spectral slopes of galaxies at $z\gtrsim7$ indicate that they may be dust-free \cite[][but see \citealt{Dunlop11}]{Bouwens10a}.  Given the uncertainty, we leave our values uncorrected at these higher redshifts.  

\begin{figure}
\begin{center}
\includegraphics[width=\columnwidth]{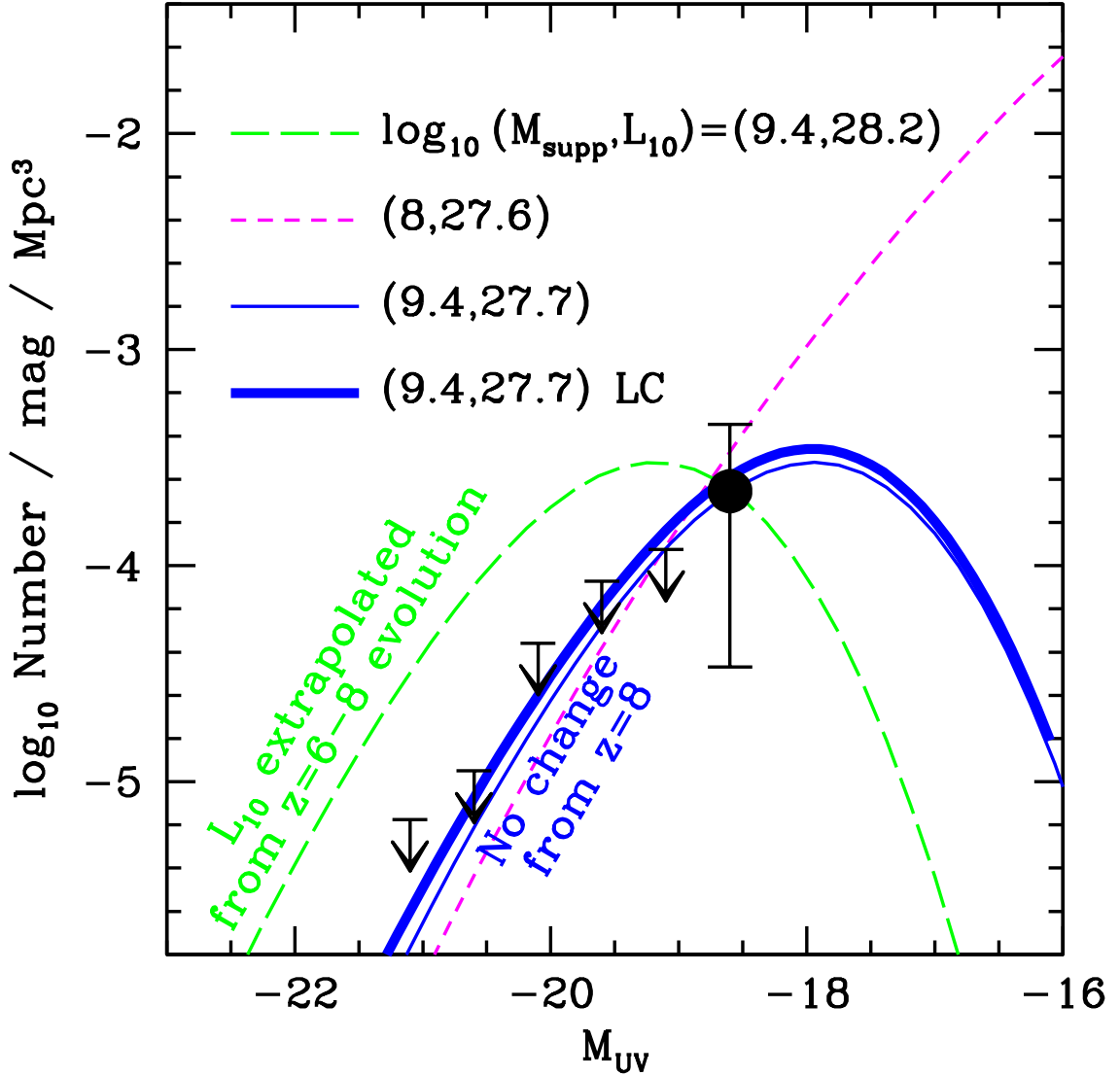}
\caption{\label{fig:LF10} The luminosity function of J-dropouts.  The point and upper limits show results from Ref. \cite{Oesch11}, while the curves show the LF generated with the ML11 model for different combinations of the free parameters $M_{\rm supp}$ and $L_{10}$.  The solid, blue and short-dashed, magenta lines have ${\rm log_{10}}\,M_{\rm supp}=9.4$ and 8, respectively, with ${\rm log_{10}}\,L_{10} \approx 27.7$.  The long-dashed, green curve assumes ${\rm log_{10}}\,(M_{\rm supp}, L_{10})=(9.4, 28.2)$ with the value of $L_{10}$ extrapolated to $z\approx10$ based on the 0.24 dex increase per redshift interval found from $z\approx6$--8 (see Fig. \ref{fig:evol_L10}).  The thick solid curve is the same as the thin one except that the evolution in the LF along the light-cone (LC) through the B11 redshift selection function has been taken into account.
}
\end{center}
\end{figure}

\begin{figure}
\begin{center}
\includegraphics[width=\columnwidth]{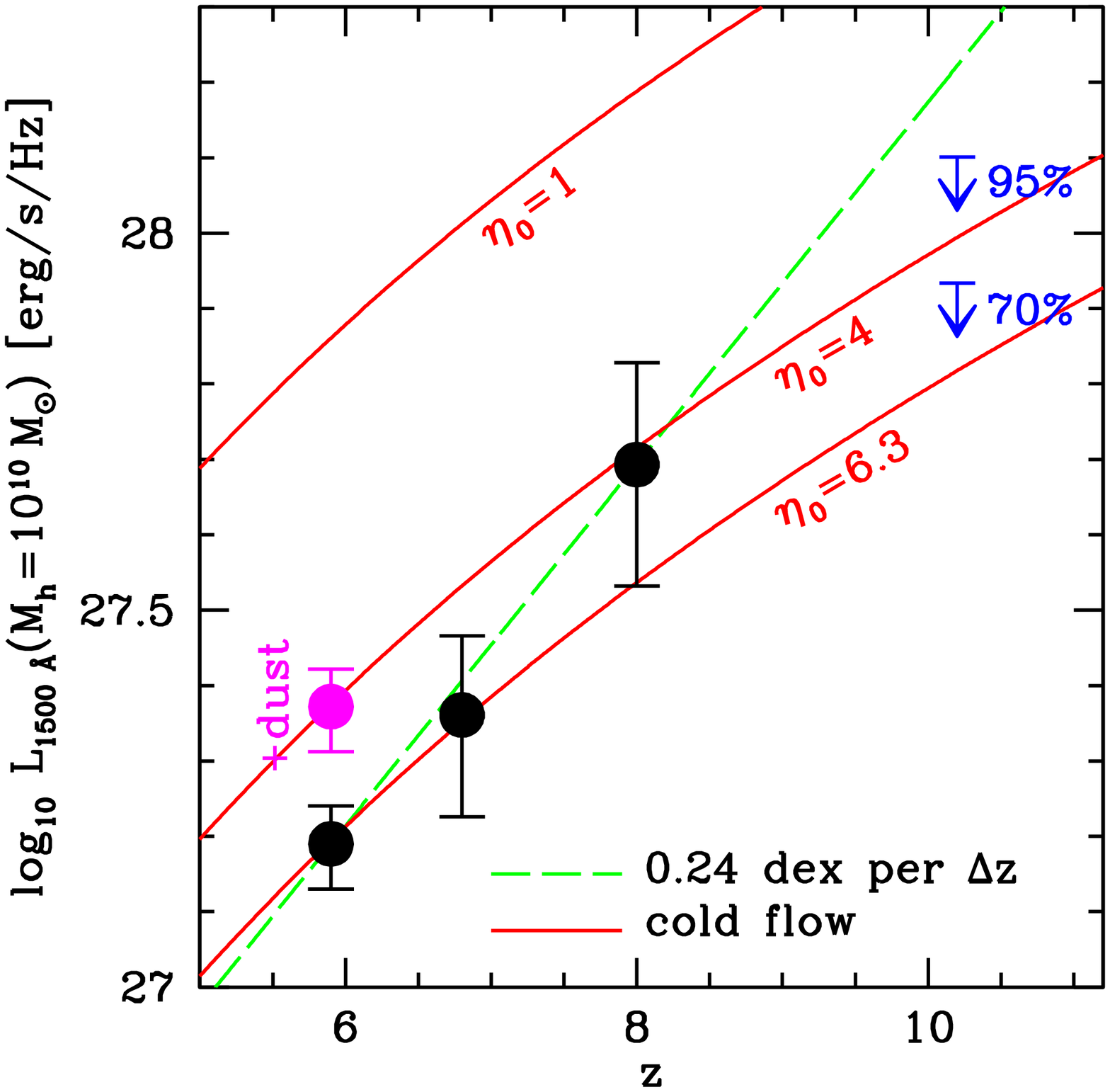}
\caption{\label{fig:evol_L10} The evolution in the UV luminosity at $10^{10}\,\msun$ from $z\sim6$--$10$.  The points show the best-fit values at $z\approx6$--$8$ with the $95\%$ confidence interval ($\approx 2\sigma$) shown at $z=6$ and $70\%$ ($\approx 1\sigma$) shown at $z\approx7$ and 8 (see footnote in main text).  The upper magenta point at $z\approx6$ includes a dust correction of 0.18 dex.  The dashed, green line marks an increase of 0.24 dex per redshift interval connecting the points at $z\approx6$ and 8, while the solid, red curves shows the predicted evolution of $L_{10}$ based on the cold-flow model of galaxy fueling and feedback in equation \ref{eq:L1500}, assuming a Salpeter IMF at solar metallicity with $\eta_0=1$, 4, and 6.3.  Blue upper limits denote the $70\%$ and $95\%$ confidence constraints placed at $z\approx10$.
}
\end{center}
\end{figure}

The single galaxy candidate observed at $z\sim10$ poorly constrains $L_{10}$ from below.  However, the lack of detections at brighter magnitudes places interesting limits from above.  Figure \ref{fig:LF10} compares the J-dropout LF with the model at $z=10.2$ for various combinations of $M_{\rm supp}$ and $L_{10}$.  The long-dashed curve shows the model LF using the $z=8.0$ value of $M_{\rm supp}=10^{9.4}\,\msun$ but extrapolating ${\rm log_{10}}\,L_{10}$ to 28.2 as might be expected if the trend of a 0.24 dex increase per redshift interval continues to $z=10.2$.  This model is clearly inconsistent with the observations, producing too many galaxies at $M_{\rm UV}=-19.6$ and perhaps too few at $M_{\rm UV}=-18.4$.  Tuning $M_{\rm supp}$ cannot reconcile such a large value of $L_{10}$ with the data.

On the other hand, the solid and short-dashed lines in Figure \ref{fig:LF10} assume no evolution in the mass-to-light ratio from $z=8.0$ and ${\rm log_{10}}\,M_{\rm supp} = 9.4$ and 8, respectively.  While ML11 showed that the larger value of $M_{\rm supp}$ is slightly preferred at $z\approx8$, neither is ruled out.  Both also match the $z\sim10$ data well.  Given that the model LFs for these values are near the two observed upper limits at $M_{\rm UV}=-18.9$ and -19.6, we conclude that the average galaxy brightness for a given halo mass cannot increase as significantly from $z\approx8$--10 as between $z\approx6$--$8$.  

Figure \ref{fig:evol_L10} shows the rise in $L_{10}$, corresponding to a drop in the mass-to-light ratio, from $z\approx6$--10.  Because $L_{10}$ at $z\sim10$ is constrained weakly from below, we plot only the upper limits.  Even correcting for dust at $z\sim6$ and ignoring dust at $z\sim8$, $L_{10}$ increases significantly between these two redshifts and the respective confidence intervals do not overlap.  The implied change in galaxy formation over this redshift range contradicts the predictions of other studies that the evolution of the LF at $z\gtrsim6$ is entirely due to the behavior of the HMF \cite[e.g.,][]{Trenti10}.  Of course, an individual galaxy does become brighter with {\it{decreasing}} redshift \cite[e.g.,][]{Finlator11}, but its halo mass does not remain fixed.  The increase in luminosity with redshift at constant halo mass is an important example of how the intrinsic properties of galaxy formation are masked by the behavior of the HMF.  The population of massive halos shrinks faster with increasing redshift than the population of small ones, thus making lower mass halos become more typical.  This effect ``wins" in determining $M^{\star}_{\rm UV}$ for a Schechter-fitted LF even though galaxies at each mass become brighter.  The observed decline of $M^{\star}_{\rm UV}$ is due in much larger part to changing halo abundances than to changes in galaxy properties.

That halos of a fixed mass host brighter galaxies on average at earlier times is expected from the increased growth rate of halos and the correspondingly larger galaxy fueling rate by cold flows.  In \S\ref{sec:galfuel}, we will quantitatively explore the consistency of this prediction.

\section{$\sigma_{\rm L}$}\label{sec:sigL}

The parameter $\sigma_{\rm L}$ describes the logarithmic spread in the luminosity of galaxies hosted by halos of the same mass.  In a scenario where the galactic luminosity is due to mergers, $\sigma_{\rm L}$ combines the rate at which galaxies grow fainter by using up gas with the amount of time since the last merger event.  For example, if the time since the last merger, $t_{\rm m}$, in a given galaxy is much less than the star formation timescale $t_{\rm sfr}$, then that galaxy's luminosity will be in the bright tail of the distribution in equation \ref{eq:LDF}.  On the other hand, if $t_{\rm m} \gg t_{\rm sfr}$, the galaxy will have used up most of its gas in the starburst and have a luminosity in the faint tail of the distribution for its host halo mass.  In a cold-flow picture for high-redshift LBGs (as further discussed in \S\ref{sec:galfuel}), $\sigma_{\rm L}\approx0.25$ is the result of the factor of a few variation in the galactic baryon accretion rate for a give halo mass \cite{Dekel09a, McBride09}.

The value of $\sigma_{\rm L}$ determines how reliably the slope of the LF reproduces that of the HMF.  For $\sigma_{\rm L}=0$ with $L_{\rm c} \propto M_{\rm halo}$, every galaxy with the same host halo mass also has the same luminosity, so the LF has exactly the same shape as the HMF (at masses sufficiently larger than $M_{\rm supp}$).  Increasing $\sigma_{\rm L}$ mixes halos of different masses into the same luminosity bin and effectively flattens the LF.  If we try to fit $\sigma_{\rm L}$ {\it{a priori}} from the data (while, for simplicity, keeping $M_{\rm supp}$ fixed), we find that the constraints from below are poor.  This is expected since the shape of the LF is not too far from that of the HMF.  However, $\sigma_{\rm L}$ is restricted to be less than $\sim 0.5$ (1), assuming ${\rm log_{10}}M_{\rm supp}\sim9.4$ (8).  Physically, this implies that the luminosities of galaxies hosted in halos of the same mass are all within about an order-of-magnitude of each other (or alternatively, within about 2.5 magnitudes of each other.

\section{Observing along the light-cone}\label{sec:lc}

In this section, we discuss the effect of a rapidly evolving LF on a J-dropout survey that samples galaxies from a large redshift range. 

The broad photometric bands used in dropout surveys of high-redshift galaxies result in significant rerdshift-error.  This owes to the uncertain location within the band at which a galaxy's spectral break falls.  In their Fig. 2, for example, B11 show the redshift selection functions of their J-dropout survey spread over a range of almost $\Delta z\approx2$.  In \cite[][hereafter ML08]{ML08b}, we demonstrated that, due to the rapid evolution of the LF, there should be far more galaxies in the observed sample at the low-redshift end of the redshift distribution than at the high-redshift end simply because the abundance of those galaxies is much higher at a later point in the history of the universe.  Consequently, the mean redshift of observed galaxies is lower than that of the photometric window.  For idealized assumptions about the LF and the shape of the broad-band filters, we found that the resulting observed abundance of galaxies in $10^{10}\,\msun$ halos could be about a factor of two higher than the true value expected at the mean photometric redshift.

Here we use the ML11 LF model, with ${\rm log_{10}}\,(M_{\rm supp}, L_{10})=(9.4, 27.7)$ (see \S\ref{sec:evol}), along with the exact redshift selection function generated by the B11 J-dropout photometric selection criteria, to more carefully consider this effect.  We calculate the difference between the abundance of galaxies at the mean selection redshift of $z=10.2$ and that in a survey along the line-of-sight light-cone taking into account the evolving LF through the photometric band.  We find that the density of observed J-dropouts with magnitude $M_{\rm UV}=-18.4$ is about $17\%$ higher than the abundance at $z=10.2$ owing to the fact that the mean redshift of the observed galaxies is actually $z=9.8$ (see Fig. \ref{fig:LF10}).  For reference, the star formation rate density required to keep the universe ionized is $12\%$ higher at $z=9.8$ than at $10.2\%$.  The effect increases with increasing brightness causing a slight flatting of the LF (for $M_{\rm UV}=-20.4$ the observed abundance is $35\%$ higher than at $z=10.2$ and the mean redshift sampled is $z=9.7$).  As discussed in ML08, we ignore the slightly varying differential volume element of the survey with redshift as a negligible contribution. 

While the effect of observing along the light-cone of the B11 survey is not negligible, it is less significant than previously anticipated and less than the error already placed on the measured abundance.  There are three important reasons for this.  (1) In ML08, we assumed a gaussian distribution for the redshift selection function in ML08 with an exponential tail trailing to lower redshifts, while the exact distribution for the B11 J-dropout survey reaches down to exactly zero\footnote{The probability that galaxies at $z<8$ are included in the sample may not be exactly zero but simply below the resolution of the B11 simulation depending on the number of mock galaxies used.  Since the LF increases exponentially with decreasing redshift, even an exponentially vanishing selection probability may create a tail of galaxies distributed to much lower redshifts.} at $z<8$.  (2) The current concordance value of $\sigma_8=0.82$ from the {\it{Wilkinson Microwave Anisotropy Probe}} 7-year results \cite{Komatsu11} is slightly higher than the 3-year value of $\sigma_8=0.776$ \cite{Spergel07} used in ML08, which results in a HMF that evolves slightly less quickly at $z\sim10$.  (3) The average mass of halos corresponding to a particular brightness in the ML11 model described in \S\ref{sec:model} is significantly smaller than in the model of Ref. \cite{SLE07} for the LBG mass-to-light ratio that we used in ML08, and so, the abundance of these halos varies less rapidly with redshift.  Not only is the average brightness of halos at a fixed mass larger in the ML11 model than in Ref. \cite{SLE07}, but the effect becomes even larger when we include the decrease in the mass-to-light ratio from $z\approx6$, where the model in Ref. \cite{SLE07} was calibrated, which was not originally included in ML08.  Moreover, the luminosity distribution function allows halos of a given mass to dominate the population of galaxies brighter than their average; despite the fact that few of these halos shine with  luminosities brighter than the average, they are much more numerous in total than are more massive halos.  These effects conspire to slow the evolution of the LF in the range over which the survey is observed.  On the other hand, if models that predict much higher halo masses for a given luminosity, such as the one in Ref. \cite{Trenti10}, are correct, then the variation in the LF over the redshift range of the survey will be substantially larger, and the effects described in ML08 may be quite significant.

\section{The evolution of galaxy fueling}\label{sec:galfuel}

Using a simple but physical model for galaxy formation, we attempt to describe the evolution of the average galaxy luminosity for a given halo mass at $z \gtrsim 6$.  If the star formation time scale is much less than the dynamical time on which galaxies grow \cite{Dave11}, then the star formation rate is a balance between the rate at which galaxies are fueled by baryons, $\dot{M}_{\rm acc}$, and rate at which large-scale galactic outflows, ubiquitous at high-redshift \cite[e.g.,][]{Steidel10}, deplete the gas available for star formation, $\dot{M}_{\rm W}$.
\begin{equation}\label{eq:sfr}
\dot{M}_{\rm SFR} = \dot{M}_{\rm acc}-\dot{M}_{\rm W}.
\end{equation}
Here, we have also assumed that a negligible fraction of gas is accreted onto a central black hole \cite{MF12}.

In the cold flow model, the infall of cold gas onto the galactic disk traces the buildup of dark matter in the halo since the gas is never shock-heated to high temperatures \cite{Keres05, Dekel09a}.  Following Ref. \cite{McBride09}, the high redshift behavior of this fueling rate is approximated by 
\begin{equation}\label{eq:Macc} 
\dot{M}_{\rm acc} \approx 3\,{\rm \msun/yr}\,\left(\frac{M_{\rm halo}}{10^{10}\,\msun}\right)^{1.127}\,\left(\frac{1+z}{7}\right)^{2.5}\,\left(\frac{f_{\rm b}}{0.16}\right),
\end{equation}
where $f_{\rm b}$ is the cosmic baryon fraction.  Equation \ref{eq:Macc} indicates that gas is accreted more quickly at higher redshifts for fixed halo mass.  

According to the momentum-driven wind model of Ref. \cite{Murray05}, the wind outflow rate is proportional to $\dot{M}_{\rm SFR}/\sigma$, where $\sigma$ is the velocity dispersion.  Since $\sigma \propto M_{\rm halo}^{1/3}\,(1+z)^{1/2}$ \cite{BL01}, winds become somewhat less important at higher redshift for fixed halo mass.  We thus define a mass-loading factor of
\begin{equation}\label{eq:eta_w} 
\eta_{\rm w}\equiv \frac{\dot{M}_{\rm w}}{\dot{M}_{\rm SFR}}=\eta_0\,\frac{100\,{\rm km/s}}{\sigma}
\end{equation}
that describes the relationship between the star formation rate and the wind outflow rate.  Here, $\eta_0$ is a free parameter defining the normalization of the wind efficiency with $\eta_0=1$ in the original derivation from Ref. \cite{Murray05}, and $V_{\infty} \sim 3\,\sigma$ is the wind velocity at infinity.  We also note that the recycling of these outflows back onto the galaxy does not affect the baryon fueling of equation \ref{eq:Macc} at these early times \cite{Oppenheimer10}.

Assuming a fixed UV luminosity (at 1500 \AA) to SFR ratio, $A \equiv L_{1500}/\dot{M}_{\star}=8\times10^{27}\,{\rm erg/s/Hz/(\msun/yr)}$ for a Salpeter IMF and solar metallicity \cite[][but see also ML11]{Madau98} and combining with equations \ref{eq:sfr}, \ref{eq:Macc}, and \ref{eq:eta_w} gives a galaxy UV luminosity as a function of mass and redshift for a choice of $\eta_0$.  
\begin{equation}\label{eq:L1500}
L_{1500}=\frac{A\,\dot{M}_{\rm acc}}{1+\eta_{\rm w}}.
\end{equation}
When $\eta_0 \gg 1$ and most of the accreted gas is expelled through winds, $L_{1500} \propto M_{\rm halo}^{\sim0.8}\,(1+z)^2$, while $L_{1500} \propto \dot{M}_{\rm acc}$ when $\eta_0 \ll 1$ and most of the accreted gas is turned into stars.  In either case, the dependence on $M_{\rm halo}$ is close to linear, as calculated in the excursion-set analysis of ML11.  In this model, $\sigma_{\rm L}\approx0.25$ results from the factor of a few spread in the galactic accretion rate from galaxy to galaxy around the mean given in equation \ref{eq:Macc} \cite{Dekel09a, McBride09}.

Figure \ref{fig:evol_L10} shows the evolution of $L_{10}$ predicted by the model of galaxy fueling and feedback represented by equation \ref{eq:L1500}.  Normalizing the model to the value of $L_{10}$ at $z\approx6$ gives $\eta_0 \approx 4.0$ and 6.3 in the dust-corrected and uncorrected cases, respectively.  A standard wind mass-loading factor with $\eta_0=1$ produces galaxies that are approximately three times too bright for their halo mass.  This super-wind requirement was previously seen in cosmological numerical simulations \cite{Dave06, OD08} which considered $\eta_0=3$ (in our notation).  Further, a more top-heavy IMF would produce a higher ratio of $L_{1500}/\dot{M}_{\star}$ and require either more dust extinction or even stronger winds to reproduce our values of $L_{10}$.

With the model normalized to the value of $L_{10}$ at $z\approx6$, the predicted evolution roughly agrees with the behavior we have extracted from the observed LFs.  This consistency is maintained out to $z\approx10$, where the a simple extrapolation of the 0.24 dex per redshift interval evolution from $z\approx6$--8 fails.  For $\eta_0=4.0$ (6.3), the model predicts a star formation rate density of $0.03\,{\rm (}0.02{\rm )}\,{\rm \msun/yr/Mpc^{3}}$ at $z=9.8$ (see \S\ref{sec:lc}) if $M_{\rm supp}=10^8\,\msun$ and $8\,{\rm (}5{\rm )}\,\times10^{-4}\,{\rm \msun/yr/Mpc^{3}}$ if $M_{\rm supp}=10^{9.7}\,\msun$.  This compares to a value of $2.5\,\times10^{-3}\,f_{\rm esc}^{-1}\,C\,{\rm \msun/yr/Mpc^{3}}$ required to maintain a reionized intergalactic medium \cite{Madau99}, where $f_{\rm esc}$ is the ionizing escape fraction and $C$ is the clumping factor.

\section{Comparison with recent numerical models}\label{sec:compare}

In this section, we compare our findings with those from several recent numerical and semi-analytic models of high-redshift galaxies, all of which produce reasonable agreement with measured LFs.  With the flexibility of our analytic framework, we can calibrate the physics from the observations rather than predict the LF {\it{a priori}}, and our resulting mass-to-light ratios consequently ``split the difference" between those from other works.

The numerical simulations of Ref. \cite{Finlator11} predict LFs that are systematically brighter than observed.  While they included some dust extinction calibrated to local measurements, excluding dust from their models only exacerbates the discrepancy.  Given the blue continuum slopes of high-redshift sources \cite{Bouwens10a}, assuming additional extinction is probably not justified even though it may bring our estimates of the mass-to-light ratio into agreement.  The shape of their conditional luminosity functions at $z=5.5$ are also quite different from ours.  While our merger-tree analysis predicts a single-peaked distribution (Eq. \ref{eq:LDF}), theirs exhibit a power-law behavior.  At this redshift, our model would indeed anticipate a rising abundance of fainter galaxies hosted by $1.7\times10^{10}\,\msun$ and $3.4\times10^{10}\,\msun$ halos at least down to $M_{\rm UV}=-18$ (i.e., the faint limit of their Fig. 14).  However, we would have expected the distribution for $1.6\times10^{11}\,\msun$ halos to have turned over by around $M_{\rm UV}=-19.5$.  The contribution from faint satellites, neglected in our treatment, may be responsible.  We note that their simulation includes momentum-driven winds with mass-loading factors similar to what we found in \S\ref{sec:galfuel}.

On the other hand, the simulations by Ref. \cite{Salvaterra11} generate galaxies at $z=7$ that are about a factor of two (i.e., about 0.7 magnitudes) more luminous than ours for $10^{10}\,\msun$ halos.  Yet, curiously, their resulting LF at this redshift is shifted somewhat from the data toward {\emph{fainter}} magnitudes.  Some of this discrepancy may result from their supernova wind prescription, which assumes an initial wind velocity of $500\,{\rm km/s}$ for all halos even though supernovae may not expand very far into the dense interstellar media of high-redshift galaxies \cite{Thompson05}.  It is difficult, however, to compare their results either to ours or to observations given their relatively small box size (only $10\,h^{-1}\,{\rm Mpc}$), which contains a very limited sample of simulated bright galaxies.  Still, their detailed treatment of metal-enrichment may make their conclusion of a negligible dust contribution to the LF fairly reliable.

Finally, Ref. \cite{Raicevic10} use a semi-analytic model of galaxy formation, combining N-body simulations of dark matter with analytic prescriptions for star formation, to model the high-redshift LF.  Their results are significantly different from both Ref. \cite{Finlator11} and Ref. \cite{Salvaterra11} in that both mergers and a topheavy IMF are required to reproduce observations.  However, it is possible that their need for a topheavy IMF is a result of over-quenching by their strong feedback prescription from supernovae-driven winds.  At the same time, their requirement for mergers may result from the long timescale for quiescent star formation assumed from low-redshift observations; if high redshift galaxies have higher molecular fractions for the same total gas mass \cite[e.g.,][]{Obreschkow09a}, star formation will proceed more efficiently that this estimate \cite[e.g.,][]{Krumholz09b}.

\section{Discussion and conclusions}\label{sec:discussion}

We have investigated the observed evolution in the LF of LBGs from $z\sim6$--10 in the context of a rapidly changing halo mass function to disentangle the corresponding variation in galaxy formation.  Our analysis improved on our previous work in ML11 by considering LFs based on the complete two-year HUDF09 data as well as incorporating information from null-detections at bright magnitudes.  

From $z\approx6$--8, we find a significant rise in $L_{10}$--the luminosity at 1500 \AA of a galaxy hosted in a halo of $10^{10}\,\msun$ and the parameter that describes the average galaxy mass-to-light ratio--which is inconsistent with the predictions of other analytic models \cite[e.g.,][]{Trenti10}.  Despite the detection of only a single J-dropout at $z\approx10$, we were also able to place interesting constraints on galaxy formation at this high redshift.  The upper limits on the number of galaxies with $M_{\rm UV} \lesssim -19$ were particularly useful in this regard.  We were able to rule out a decrease in the average mass-to-light ratio of galaxies from $z\sim8$--$10$ comparable to the evolution from $z\sim6$--$8$.  This result would remain unchanged if the single B11 galaxy candidate were actually a spurious detection or a low-redshift interloper (B11 estimate a $20\%$ total probability for such an event).  However, there are two observational effects that could allow for a higher halo luminosity (i.e. lower mass-to-light ratio) at $z\sim10$.  First, the strict set of selection criteria designed to minimize low-redshift interlopers could have resulted in undetected sources that should be considered in the LF \cite[see, e.g.,][]{McLure11}.  Second, light outside the selection aperture could have been missed due to surface brightness dimming; the B11 object and upper limits were not corrected for this potential missing signal.  The correction can increase the brightness by $\sim 0.5$ magnitudes at $z\approx8$ \cite{Bouwens11b}.  Since the surface brightness of a galaxy with fixed size and spectral luminosity decreases\footnote{While the Tolman scaling of bolometric surface brightness with redshift is ${\rm SB} \propto (1+z)^{-4}$ \cite{Tolman1930, LS01}, the spectral surface brightness of fixed emitted $L_{\nu_{\rm em}}=dL/d{\nu_{\rm em}}$ is multiplied by an extra factor of $(1+z)$ to account for the increased observed ``bandwidth" at higher redshift.} as $(1+z)^{-3}$ and assuming galaxy size scales as $(1+z)^{-0.875}$ \cite{WL11}, we estimate that the correction at $z\approx10$ could be as large as a full magnitude making the data not inconsistent with the 0.24 dex per redshift trend shown in Figure~\ref{fig:evol_L10}.  Additionally, if we have failed to account for a feedback process that shuts off star formation and creates an ``effective duty-cycle," reducing the abundance of halos seen as galaxies, then the true values of $L_{10}$ will be higher than we have calculated with the degree of underestimation depending on the mass, luminosity, and redshift dependencies of the feedback mechanism.

While we calculated the above results with $\sigma_{\rm L}=0.25$ based on the ML11 merger tree calculation, we also examined to what extent this value is independently constrained by the data if left to vary.  An increase in $\sigma_{\rm L}$ flattens the slope of the LF by mixing halos from a larger mass range into the same luminosity bin.  We find that the data limit the logarithmic spread in galactic luminosity at fixed halo mass to be less than about an order-of-magnitude.

We then presented a simple but physical model for galaxy fueling from cold-flows and feedback from momentum-driven winds to explain the evolution we found in $L_{10}$.  Normalizing to $L_{10}$ at $z \approx 6$, this model does an excellent job of predicting the behavior of the average galaxy luminosity out to $z \approx 10$.  Thus, it appears that, on average, galaxy formation at these redshifts can be described as a steady-state balance among gas accretion, star formation, and wind ejection, as discussed in \cite{Dave11}, while ignoring the growth of the gas reservoir in the disk.  Moreover, if galaxy formation proceeds in halos as small as $10^{8}\,\msun$ \cite{BL01}, then the star formation rate density at $z\approx10$ should be high enough to keep the universe ionized as long as $f_{\rm esc}^{-1}\,C\lesssim10$.  

However, future data from WFC3, both deeper and in multiple fields (to reduce cosmic and Poisson variance), will improve constraints on the observed LF and further test these predictions even before the {\it{James Webb Space Telescope}} comes online.

\section{Acknowledgements}

We thank Avi Loeb, Steve Furlanetto, Molly Peeples, and Kristian Finlator for useful discussions and Rychard Bouwens for supplying his redshift selection functions for HUDF09 J-dropouts.  We further acknowledge helpful notes from the JCAP referee.

\bibliography{evolLF_JCAP_rev2.bbl}

\end{document}